# Clustering online social network communities using genetic algorithms

Mustafa H. Hajeer[*]    Alka Singh[*]    Dipankar Dasgupta[*]    Sugata Sanyal[#]

[*]Center for Information Assurance
Department of Computer Science
The University of Memphis
Memphis, TN 38138, USA

[#]School of Technology & Computer Science
Tata Institute of Fundamental Research
Homi Bhabha Road,
Mumbai-400005, INDIA

**ABSTRACT**
To analyze the activities in an Online Social Network (OSN), we introduce the concept of "*Node of Attraction*" (NoA) which represents the most active node in a network community. This NoA is identified as the origin/initiator of a post/communication which attracted other nodes and formed a cluster at any point in time. In this research, a genetic algorithm (GA) is used as a data mining method where the main objective is to determine clusters of network communities in a given OSN dataset. This approach is efficient in handling different type of discussion topics in our studied OSN –comments, emails, chat sessions, etc. and can form clusters according to one or more topics. We believe that this work can be useful in finding the source for spread of rumors/misinformation, help law enforcement in resource allocation in crowd management, etc. The paper presents this GA-based clustering of online interactions and reports some results of experiments with real-world data and demonstrates the performance of proposed approach.

## 1. INTRODUCTION

Blogging, twitting and online social networking have become an integral part of modern-day life. Studies have shown that people spend a lot of their time in online media such as Facebook, MySpace, Twitter, YouTube, Google+ and in other blogosphere. These web-based services/applications are designed to assist people and businesses alike to stay in touch, communicate and collaborate more effectively, and socialize across borders and in general, maintain a second avatar in cyberspace. The way blog sites (nodes) attract people for participating in discussion, share information and/or spread rumor is very interesting and fascinating to study. The dynamic interaction among these nodes and the period of increased interaction and manner in which such interaction get decreased can provide useful information about the current events, topics of discussion, and human behavior. In the analysis of online social network (OSN) data, the basic problem lies in the detection of groups of closely connected nodes which are called 'network communities'. In this work, we consider the relation among the nodes in OSN, and group them according to the strength of their relations/interactions. The work is motivated by the need to understand how the interactions in virtual-world can manifest real-world security concerns or law and order situations. Such a clustering of online social communities may also help to detect insider threats, employee behavior, or recognizing competitor interests in social activities as possible motivators.

In this research, a genetic algorithm (GA) is used as a data mining method where the main objective of clustering is to find network communities from a given OSN dataset. There exist some works which addresses different aspects of online social behavior and used un-weighted social network data for clustering [2], i.e. every edge in the network has equal value/weight and distributed community detection in delay tolerant networks [3, 9-10].

## 2. GA-BASED APPROACH

An OSN community can be defined as a set of users who are frequently interacting with each other and participating in some discussion e.g. group, subgroup, module, and cluster.

A Genetic algorithm (GA) is an optimization method which uses biological process as a model to find solutions in the search problem. We used a GA here for finding the clusters in OSN data having multiple values for each edge indicating individual node participating in more than one discussion groups or activities. The goal is to improve the processing and efficient interpretation of real-world OSN data, relevant knowledge, and subsequent information processing and representation. The underlying approach tries to find dense clusters using an edge removal strategy in an intelligent way based on the context of discussions.

The reasons for using GA-based approach are

- Huge search space and non-linearity in search space.
- The dynamic nature of network topology and the need of adaptive multi-fitness function.
- GAs perform global search in problem space
- GAs are easy to interface to existing simulations and models
- GAs are extensible and easy to hybridize
- GAs are remarkably noise tolerant

The implementation details of the GA approach for OSN data clustering is explained in the next section, the section 4 provides some experiments and results. The last section gives some concluding remarks.

### 2.1 Node of Attraction:

As we know, nodes in a social network are not physical nodes rather web sites, postings, user blogs, pages with some news, etc. For example, the documentary on

Invisible Children (Kony 2012) was posted at a video-sharing website, *YouTube.com* on 5th March 2012. In two weeks time, over 78 million viewers visited the site; another website *vimeo.com* also posted the same video and attracted over 16.6 million viewers during the same time. This indicates that the sites which hosted the video attracted significant number of viewers forming clusters with the posted page as the focal point.

We introduced the concept of *Node of Attraction* (NoA) which represents the most active node in a network community. This NoA is identified as the origin of a post/communication which attracts other nodes to form a cluster. The need of finding such nodes and to keep track of these nodes in social networks has many real-world applications such as detection of spreading news or rumor in the society.

## 3. IMPLEMENTATION DETAILS

In this work, a genetic algorithm is used to find clusters/groups in online network data by removing minimum number of edges while the clusters have maximum number of ties.

### 3.1 Encoding schemes
We have used two encoding schemes where each chromosome represents groups or clusters.

In the first scheme, each chromosome is variable in size, composed of edge list as shown in Figure 1 (top left). These edges are derived from nodes in each row of the table (right). The genetic process will remove edges from the chromosome to form clusters as illustrated in Figure 1 (bottom left).

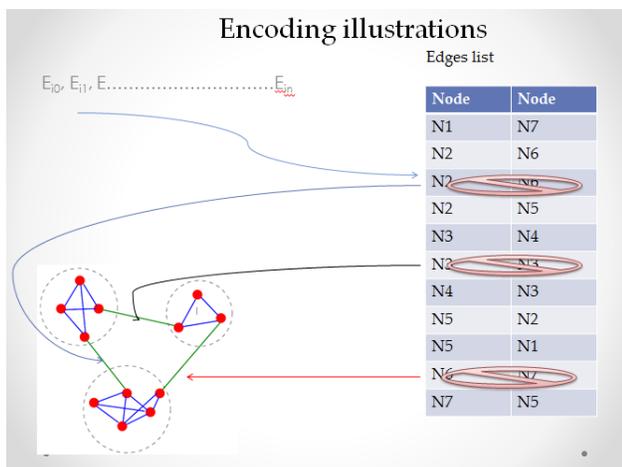

**Figure 1: First encoding scheme showing edge removal from the OSN data**

The second encoding scheme uses a real-valued representation, where the first number denotes the number of groups and the rest fields denote the points of separation between nodes to form groups as shown in Figure 2.

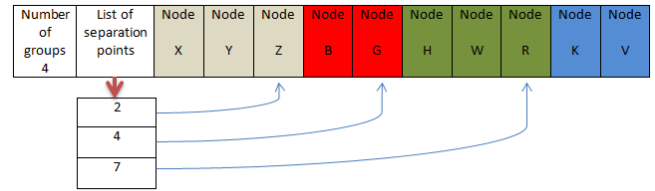

**Figure 2: The second encoding scheme**

### 3.2 The fitness function
Two properties of clustering such as high intra-cluster similarity and low inter-cluster similarity are considered for fitness measure of each chromosome. Accordingly, the number of edges connected to a node form a group are considered for fitness, where we calculate the clossness of groups/clusters and penalize the chromosome with smaller groups with same fitness. The fitness function uses the most updated edge values rather than old values in order to adapt to the changes in the interactions.

### 3.3 Setting of GA control parameters
To create diversity in the candidate solution list, the initial population is generated with random edges, and the fitness for each chromosome is calculated.

We run a steady-state GA with two different datasets. A single-point crossover with probability of 0.85, population size 100, maximum evaluations 10,000, and the binary tournament selection operator are used. The high rate of mutation is used so that the genetic search can adapt to dynamic user interactions (as reflected in data) over time. After crossover and mutations are performed, each chromosome is repaired (if necessary) in order to remove repeated edges in the chromosome.

For the second encoding scheme, we used a different type of crossover where each chromosome uses the swap operator where some genes are exchanged, and the mutations are restricted to the second and first parts of the chromosome.

## 4. EXPERIMENTS AND RESULTS

In order to determine the correct NoA for an attribute, the chromosome exhibiting the best fitness is searched and then it is decoded/ mapped to form groups. After the mapping operation, these groups are sent to a function which finds the NoA (the node with maximum number of edges) and is saved in a file to keep track of all NOAs over a period of time.

To demonstrate the performance of the GA-based clustering approach, we first experiment with a hand-crafted synthetic datasets and then we use some real-world OSN data.

### 4.1 Synthetic (small) dataset
Table 1 shows the small dataset containing 15 edges each with 3 attributes as different type of interactions between a pair of nodes. Such interactions represent multi-valued edges among the nodes. Here the interactions/edges represent number of emails, number of comments and the number of posts between nodes. Figure 3 is the corresponding network diagram (of Table 1) with some multi-valued edge. If there is no edge between any two nodes which indicates that attribute values between these two nodes are zero.

Table 1: A small synthetic dataset

| Node A | Node B | Number of emails | Number of posts | Number of comments |
|---|---|---|---|---|
| 1 | 2 | 4 | 4 | 4 |
| 1 | 3 | 3 | 3 | 3 |
| 1 | 4 | 4 | 4 | 4 |
| 1 | 5 | 4 | 5 | 5 |
| 5 | 4 | 3 | 4 | 4 |
| 5 | 3 | 4 | 3 | 3 |
| 2 | 3 | 3 | 3 | 3 |
| 2 | 4 | 3 | 3 | 3 |
| 4 | 7 | 1 | 29 | 1 |
| 8 | 14 | 2 | 1 | 30 |
| 5 | 6 | 1 | 39 | 1 |
| 6 | 7 | 5 | 4 | 3 |
| 6 | 8 | 3 | 5 | 4 |
| 6 | 9 | 4 | 4 | 4 |
| 7 | 8 | 4 | 3 | 5 |
| 7 | 9 | 3 | 4 | 4 |
| 8 | 9 | 4 | 5 | 5 |
| 6 | 10 | 1 | 1 | 35 |
| 10 | 11 | 4 | 4 | 3 |
| 10 | 12 | 3 | 3 | 3 |
| 10 | 14 | 4 | 3 | 4 |
| 11 | 12 | 2 | 4 | 5 |
| 11 | 13 | 3 | 3 | 4 |
| 11 | 14 | 5 | 4 | 5 |
| 12 | 13 | 4 | 5 | 4 |
| 13 | 14 | 3 | 4 | 5 |
| 12 | 14 | 4 | 3 | 4 |
| 14 | 15 | 3 | 4 | 5 |

After running the program, we observed that the GA is able to form clusters corresponding to each discussion groups (attributes) as shown in Figure 4.

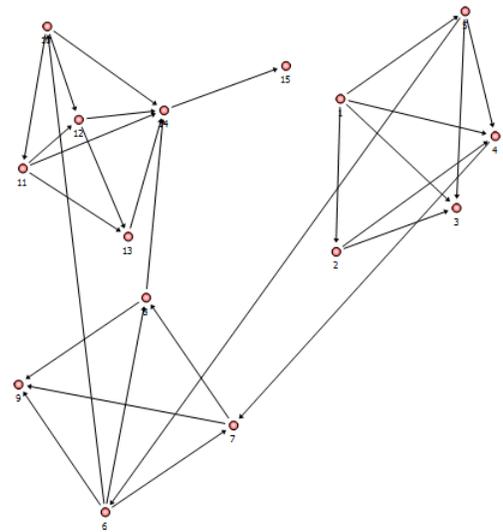

Figure 3: A graphical representation of the small dataset.

Figure 4(a) graphically represents results of a GA run which contains edges list forming network clusters according to the first attribute of the synthetic dataset (Table 1).

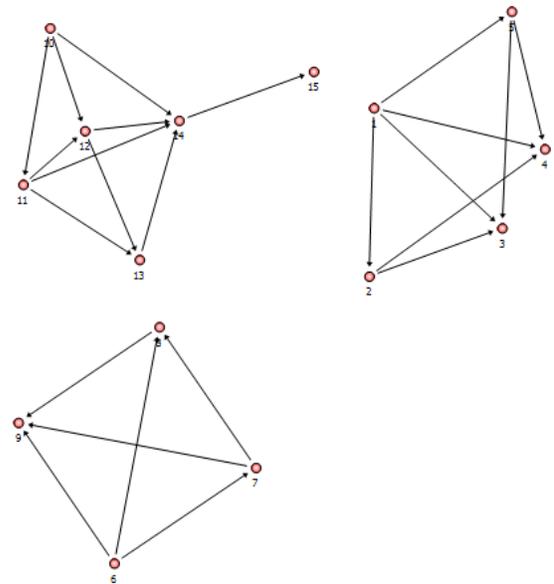

Figure 4(a): GA-based clustering according to number of Emails.

Similarly, Figure 4(b) and Figure 4(c) show the clusters formed based on other two attributes: number of comments and number of posts.

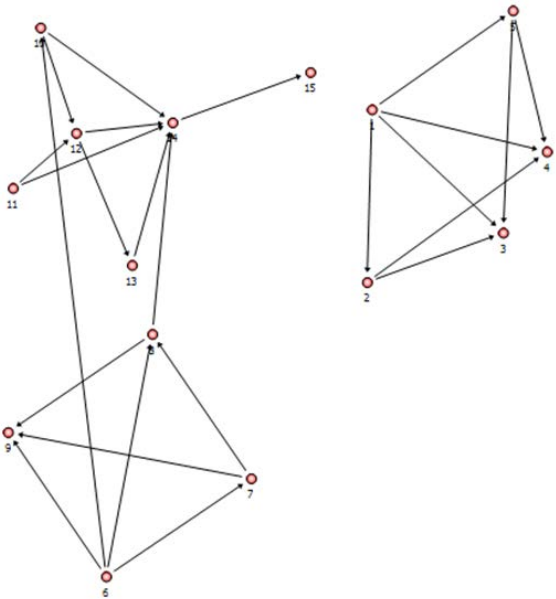

**Figure 4(b): Clustering according to number of comments.**

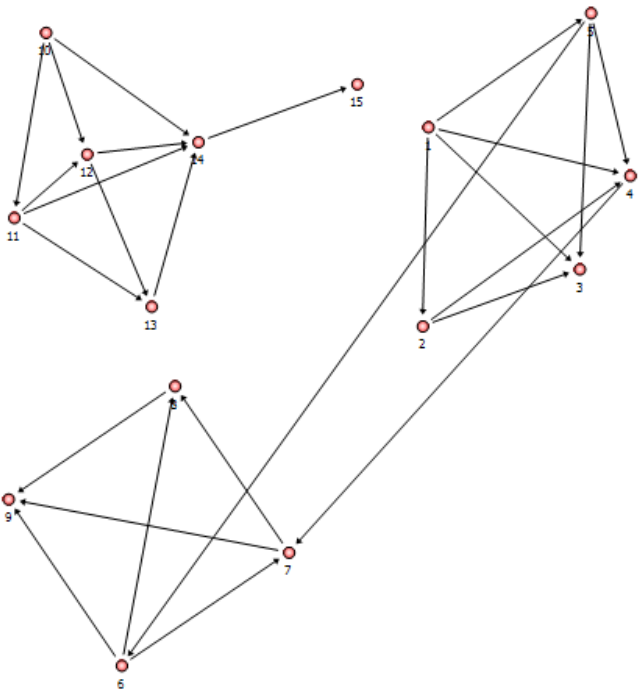

**Figure 4(c): Clustering according to number of posts.**

As per definition, the NoA for different attribute can be observed in each figure above. Accordingly, in figure 4(a) NoAs are {1, 9 and 14}, in figure 4(b) NoA nodes are {14 and 1} and in figure 4(c) these nodes are {6 and 14}. However, this provides a simplest analysis of real online social network interactions. Only one attribute is not sufficient to classify the entire social network. In particular, groups should not be clustered only on the basis of email exchanged among nodes. Other attributes should be selected in conjunction.

Now, let us take a look at figure 5(a) and Figure 5(b) below, where inter-group interactions and overlapping are demonstrated. If the OSN data are clustered according the first and the second attributes as per the dataset (Table 1), the overlapped groups are formed.

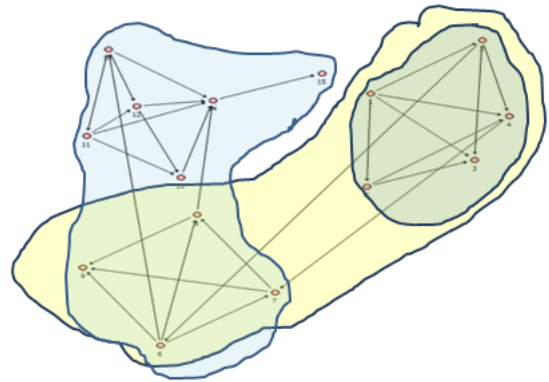

**Figure 5 (a): Groups formed according multi-valued edges**

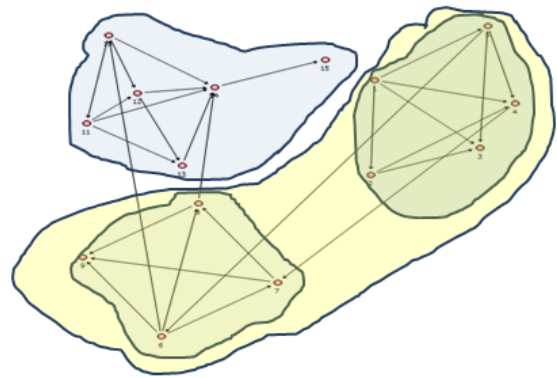

**Figure 5(b): Groups formed according multi-valued edges**

If we consider multiple attributes, it is observed that some nodes connect two or more separate clusters; such a node is identified as the "linkage node". These nodes help to analyze the interactions between two different discussion groups or topics (figure 5).

To illustrate the dynamic interactions in social networks and to demonstrate the performance of our GA-based clustering approach, we changed some attribute values during the run. At time $T_0$ during the execution of the algorithm, clusters started to form, based on the dataset as shown in figure 6, and after that some changes are made in the first attribute. This process is illustrated through the following example:

Here the second group is formed with the nodes {6, 7, 8, 9} before changes were made and NoA was the node number {9}. Sometime after $T_0$, node X is added to the second group and it was connected to all group members. Since it was connected with all group members, and it was observed that the system immediately considers X as the new node of attraction.

So after initial clustering, the groups' formation adapts to changes made to the network interaction. Table 2 shows the results of these dynamic changes, where the changes take place between time period $T_0$ and $T_1$ as follows:

- Node X was added to the data set.
- Edges were connected from X to nodes {6,7,8,9}

and the changes which take place between time $T_1$ and $T_2$ are:

- Node Y added to the dataset.
- Edge from X to node 6 and node 7 is removed to reduce its value.
- Edges from Y added to nodes {6, 7, 8, 9, X}. Hence node Y becomes the new node of attraction.

**Table2: second group dynamics**

| Time | Group members | Node of attraction |
|---|---|---|
| $T_0$ | 6,7,8,9 | 9 |
| $T_1$ | 6,7,8,9,X | X |
| $T_2$ | 6,7,8,9,X,Y | Y |

Figure 6 illustrates these changes where the blue edges represent the regular nodes which are added to the dataset and red edges show new Node of Attraction.

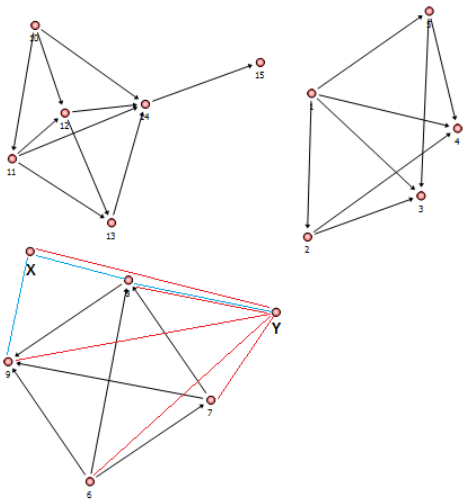

**Figure 6: Cluster results with dynamic changes in the synthetic dataset**

### 4.2 Experiments with real-world dataset

The real-world dataset considered here is taken from Stanford Large Network Dataset Collection [5], it is called Gnutella peer to peer network. This dataset contains 20,777 edges and random value is assigned to each edge (Figure 7),

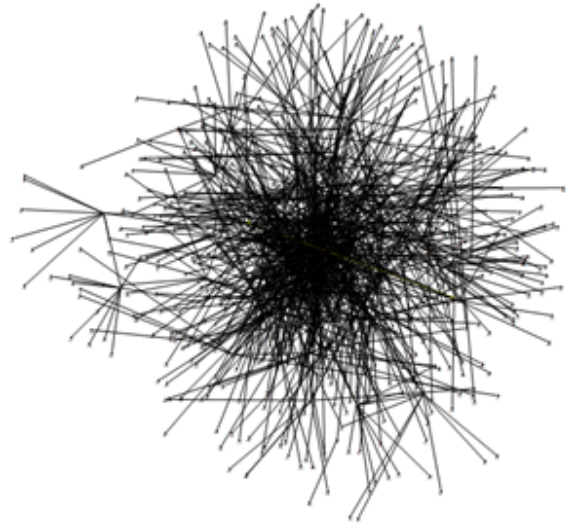

**Figure 7: An example network dataset with 1000 edges [5]**

Our solutions contain the groups which start to form slowly with smaller groups with strong ties. Hence after each (100) iterations we extracted the solution and observed more groups being formed in the solution set. The GA-based system responds slowly especially in the beginning (takeoff time) because of the huge solution space, but after forming the initial groups, it could quickly reflect changes to the groups, and produce better and reliable results as the time passes.

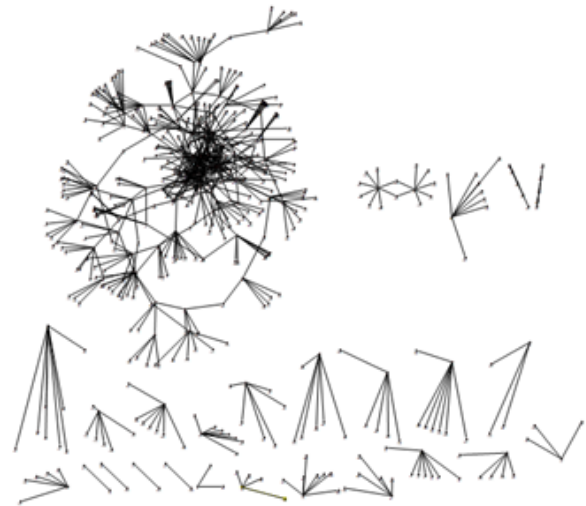

**Figure 8: Clusters formed in 1000 edges network data after running GA for 1000 iterations**

The clustering results obtained after 1000 iteration of a GA run is visualized in Figure 8. It shows that some small groups and a big group are formed. If we keep running the program, the big group would have further divided into smaller groups. It was observed that the groups which are already formed are adaptable to any change in the input dataset.

The differences between Figure 7 and Figure 8 are clearly visible; the original data representation in Figure 7 contains all nodes which are connected to each other in some manner (these figures are a presentation of clusters). The size of the groups varies based on the complexity of interactions.

Again, to illustrate the dynamism, some nodes are added to the dataset (Blue); the nodes with high value and high connectivity (shown in red) are the new nodes of attraction (Red) which are shown in Figure 9 below. It shows that updated clusters are formed in the social network interactions.

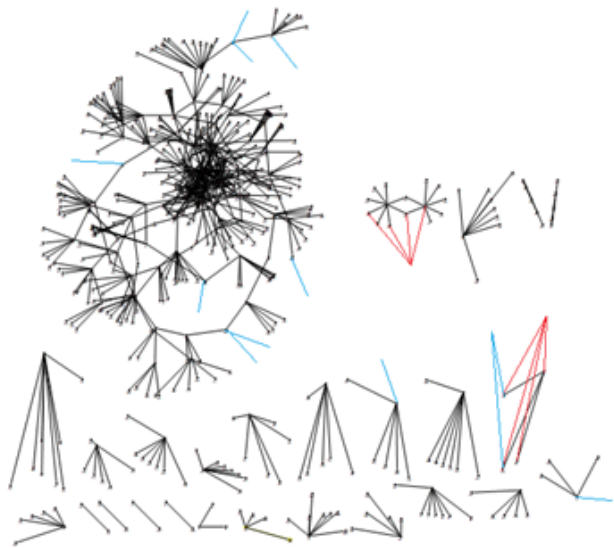

**Figure 9: Changes on the dataset during running the system causes restructuring of groups or clusters.**

Because of multi-attribute clustering technique used, this work provides the power to observe the points of connections nodes, indicating the nodes that connect more than one group, and it's considered as linkage nodes.

## 5. DATA AVAILABILITY

In the first test case (Table 1 -Synthetic dataset), we assumed that the user has access to data that have multiple weights for each connection, like weight for messages exchange and weights for discussions, but in real-world we may not have such information. So for the second set of experiments (real-world dataset), we considered equal values for all connections (i.e. either there is a connection or no connection) since we have only edge-list in the datasets [5].

## 6. CONCLUDING REMARKS

An OSN community can be defined as a set of users who are frequently interacting with each other and participate in some discussions. Determining such communities in an OSN has a wide verity of applications, such as

- Understanding the interactions among group of people
- Visualizing and navigating huge networks of NoAs
- Forming a basis for other tasks such as data mining
- Marketing, handling law and order situations

This approach allows us to form clusters based on interactions among group members in OSNs; we introduced the term NoA (Node of Attraction) which represents a node that attracts most of other nodes in the same group at a given time. The Node of Attraction in a social network, captures interaction dynamics "subsets of actors among whom there are relatively strong, direct, intense, frequent or positive ties"[1]. This NoA can help to predict the forming and merging of groups and subgroups. For example, if nodes from one sub group starts to interact with NoA of another sub group, there is a high probability that these sub groups may merge soon to form a larger group. Another benefit of NoA is that it helps to identify the source of a post/news and also may help in community detection in blogsphere.

.Because of the huge size of the solution space, we chose to use a GA approach to solve this problem, the main idea is to cluster the whole network into groups and at the same time keep track of NoAs in dynamic interactions.

The experimental run takes some time in producing reliable results but results seems to be persistent even with the dynamic changes. In case of a snapshot of the best individual (best so far) solution is desired from the population, the software is capable of producing such a result at any given time.